\def\mn{{Mon.\@ Not.\@ Roy.\@ Ast.\@ Soc.\ }}
\def \jetpl {JETP Lett.\ }
\documentclass[prd,aps,floats,preprintnumbers,preprint]{revtex4}

\usepackage{graphicx}
 
\newcommand{\be}{\begin{equation}}
\newcommand{\ee}{\end{equation}}
\newcommand{\bea}{\begin{eqnarray}}
\newcommand{\eea}{\end{eqnarray}}

\begin{document}

\preprint{YITP-09-108}
\preprint{KUNS-2247}
%\preprint{RESCUE-28-09}

\setlength{\unitlength}{1mm}

\title{Can the cosmological constant be mimicked by smooth large-scale inhomogeneities for more than one observable?}

\author{Antonio Enea Romano}
\affiliation{Yukawa Institute for Theoretical Physics and Physics Department , Kyoto University,
Kyoto 606-8502, Japan}

%\affiliation{
%Research Center for the Early Universe (RESCEU),
%Graduate School of Science, The University of Tokyo, Tokyo 113-0033, Japan}

\begin{abstract}
As an alternative to dark energy it has been suggested that we may be at the center of an inhomogeneous isotropic universe described by a Lemaitre-Tolman-Bondi (LTB) solution of Einstein's field equations.
In order to test such an hypothesis we calculate the low redshift expansion of the luminosity distance $D_L(z)$ and the redshift spherical shell mass density $mn(z)$ for a central observer in a LTB space without cosmological constant and show how they cannot fit the observations implied by a $\Lambda CDM $ model if the conditions to avoid a weak central singularity are imposed, i.e. if the matter distribution is smooth everywhere. 
Our conclusions are valid for any value of the cosmological constant, not only for $\Omega_{\Lambda}>1/3$ as implied by previous proofs that $q^{app}_0$ has to be positive in a smooth LTB space, based on considering only the luminosity distance.  

The observational signatures of smooth LTB matter dominated models are fundamentally different from the ones of $\Lambda CDM $ models not only because it is not possible to reproduce a negative apparent central deceleration $q^{app}_0$, but because of deeper differences in their space-time geometry which make impossible the inversion problem when more than one observable is considered, and emerge at any redshift, not only for $z=0$.

\end{abstract}

\maketitle
\section{Introduction}
High redshift luminosity distance measurements \cite{Perlmutter:1999np,Riess:1998cb,Tonry:2003zg,Knop:2003iy,Barris:2003dq,Riess:2004nr} and
the WMAP measurement \cite{WMAP2003,Spergel:2006hy} of cosmic
microwave background (CMB) interpreted in the context of standard 
FLRW cosmological models have strongly disfavored a matter dominated universe,
 and strongly supported a dominant dark energy component, giving rise
to a positive cosmological acceleration.
As an alternative to dark energy, it has been 
proposed \cite{Nambu:2005zn,Kai:2006ws}
 that we may be at the center of an inhomogeneous isotropic universe 
described by a Lemaitre-Tolman-Bondi (LTB) solution of Einstein's field 
equations, where spatial averaging over one expanding and one contracting 
region is producing a positive averaged acceleration $a_D$.
Another more general approach to map luminosity distance as a function of
 redshift $D_L(z)$ to LTB models has been recently
 proposed \cite{Chung:2006xh,Yoo:2008su},
 showing that an inversion method can be applied successfully to 
reproduce the observed $D_L(z)$.   

The main point is that the luminosity distance is in general sensitive 
to the geometry of the space through which photons are propagating along
null geodesics, and therefore arranging appropriately the geometry of 
a given cosmological model it is possible to reproduce a given $D_L(z)$.
For FLRW models this corresponds to the determination of $\Omega_{\Lambda}$
 and $\Omega_m$ and for LTB models it allows to determine the 
functions $E(r),M(r),t_b(r)$.

Another observable which could be used to constrain LTB models is the redshift spherical shell mass $mn(z)$ \cite{Mustapha:1997xb}, recently calculated \cite{Romano:2009qx} for a central observer up to the fifth order, which is the product of the number of sources $n(z)$ times their mass $m(z)$. This observable can be related to observations by using the following relation with the total redshift rest mass $M_T(z)$ , which is defined as the total rest mass of all the sources with redshift equal or less than $z$  :

\bea
M_T(z)=\int^z_0 4\pi mn(z')d z' ,\\
4\pi mn(z)=\frac{M_T(z)}{dz} .
\eea

%Alternatively if we are interested in a more direct relation of $mn(z)$ to observations, without having to use $M_T(z)$, $E_{RSS}(z)$ is the quantity which should be naturally %considered, since the uncertainty in the redshift determination would always imply the necessity of same integration in redshift space of $mn(z)$.

%The relation between $mn(z)$ and $E_{RSS}(z)$ given in eq.(1,2) is particularly useful for observational purposes, since an appropriate choice of the times scale $\Delta t$ can significantly avoid unwanted astrophysical evolution effects on the sources number counts $n(z)$, which is exactly the reason why $E_{RSS}(z)$ it is defined in that way.

Once an actual accurate estimation of $mn(z)$ based on observational data from galaxy surveys will be available, this quantity could be used to distinguish without ambiguity between LTB and $\Lambda CDM$ models, but an even better candidate as a discriminating observable would be the redshift spherical energy $E_{RSS}$ introduced in \cite{Romano:2007zz}, which  has the additional advantage of avoiding source evolution effects.

$E_{RSS}$ is in fact obtained by integrating $mn(z)$ over varying redshift intervals $\Delta Z(z)$ 

\bea
E_{RSS}(z)=\int \limits_{z}^{z+\Delta Z(z)} {4\pi mn(z')d\,z'}\\
t(z)-t(z+\Delta Z(z))=\Delta t
\eea

which preserve the constant time interval $\Delta t$. By choosing $\Delta t$ to be sufficiently smaller than the time scale over which astrophysical evolution is important, we can avoid the effects of the source evolution.

In this paper calculate the low redshift expansion of $mn(z)$ and $D_L(z)$ for flat $\Lambda CDM$ and matter dominated $LTB$. We then show how, if  the conditions to avoid a central weak singularity are imposed, it is impossible to mimick dark energy with a LTB model without cosmological constant for both these observables, giving a general proof of the impossibility to give a local solution of the inversion problem for a smooth LTB model.
This central singularity is rather mild, and is associated to linear terms in the energy density which lead to a divergence of the second derivative, so non smooth LTB models could still be viable cosmological models. It can be shown that the inversion problem \cite{Romano:2009mr} can be solved if the smoothness conditions we are imposing are relaxed.
This implies that the numerical solutions of the inversion problem which have been recently proposed \cite{Celerier:2009sv,Kolb:2009hn}
 must contain such a weak central singularity.
  
Our proof of the impossibility of the local solution of the inversion problem is general, since we don't use any special ansatz for the functions defining LTB models, and it does not depend on the particular value of $\Omega_{\Lambda}$.
In this regard we obtain a much stronger result than the one already well known fact that $q^{app}_0$ cannot be negative for a smooth LTB model.
In fact even if $q^{app}_0>0$, which for a flat $\Lambda CDM$ would imply $\Omega_{\Lambda}<\frac{1}{3}$, it would still be impossible to mimick the cosmological constant for both $mn(z)$ and $D_L(z)$. 

This means that the difference between a smooth spherically symmetric pressureless matter dominated Universe and a homogeneous Universe with a cosmological constant goes beyond the sign of $q^{app}_0$, and corresponds to distinctive observational features not only at $z=0$. 
From a geometrical point of view we can interpret our results as an example in which redshift space observations allow to distinguish between different cosmological models despite the fact that they don't allow us to directly probe the actual local geometry of the Universe but only to access some "compressed" information about it.

\section{Lemaitre-Tolman-Bondi (LTB) Solution\label{ltb}}
Lemaitre-Tolman-Bondi  solution can be
 written as \cite{Lemaitre:1933qe,Tolman:1934za,Bondi:1947av}
\begin{eqnarray}
\label{eq1} %
ds^2 = -dt^2  + \frac{\left(R,_{r}\right)^2 dr^2}{1 + 2\,E}+R^2
d\Omega^2 \, ,
\end{eqnarray}
where $R$ is a function of the time coordinate $t$ and the radial
coordinate $r$, $R=R(t,r)$, $E$ is an arbitrary function of $r$, $E=E(r)$
and $R,_{r}=\partial R/\partial r$.

Einstein's equations give
\begin{eqnarray}
\label{eq2} \left({\frac{\dot{R}}{R}}\right)^2&=&\frac{2
E(r)}{R^2}+\frac{2M(r)}{R^3} \, , \\
\label{eq3} \rho(t,r)&=&\frac{2 M,_{r}}{R^2 R,_{r}} \, ,
\end{eqnarray}
with $M=M(r)$ being an arbitrary function of $r$ and the dot denoting
the partial derivative with respect to $t$, $\dot{R}=\partial R(t,r)/\partial t$.
 The solution of Eq.\ (\ref{eq2}) can be expressed parametrically 
in terms of a time variable $\tau=\int^t dt'/R(t',r) \,$ as
\begin{eqnarray}
\label{eq4} Y(\tau ,r) &=& \frac{M(r)}{- 2 E(r)}
     \left[ 1 - \cos \left(\sqrt{-2 E(r)} \tau \right) \right] \, ,\\
\label{eq5} t(\tau ,r) &=& \frac{M(r)}{- 2 E(r)}
     \left[ \tau -\frac{1}{\sqrt{-2 E(r)} } \sin \left(\sqrt{-2 E(r)}
     \tau \right) \right] + t_{b}(r) \, ,
\end{eqnarray}
where  $Y$ has been introduced to make clear the distinction
 between the two functions $R(t,r)$ and $Y(\tau,r)$
 which are trivially related by 
\begin{equation}
R(t(\tau,r))=Y(\tau,r) \, ,
\label{Rtilde}
\end{equation}
and $t_{b}(r)$ is another arbitrary function of $r$, called the bang function,
which corresponds to the fact that big-bang/crunches can happen at different
times. This inhomogeneity of the location of the singularities is one of
the origins of the possible causal separation \cite{Romano:2006yc} between 
the central observer and the spatially averaged region for models
 with positive $a_D$.

%We also observe that this solution is also valid of positive $E$, if we continue it analytically choosing the branch %given by:

%\be
%sin(sqrt{-E})=
%\ee

We introduce the variables
\begin{equation}
 A(t,r)=\frac{R(t,r)}{r},\quad k(r)=-\frac{2E(r)}{r^2},\quad
  \rho_0(r)=\frac{6M(r)}{r^3} \, ,
\end{equation}
so that  Eq.\ (\ref{eq1}) and the Einstein equations
(\ref{eq2}) and (\ref{eq3}) are written in a form 
similar to those for FLRW models,
\begin{equation}
\label{eq6} ds^2 =
-dt^2+A^2\left[\left(1+\frac{A,_{r}r}{A}\right)^2
    \frac{dr^2}{1-k(r)r^2}+r^2d\Omega_2^2\right] \, ,
\end{equation}
\begin{eqnarray}
\label{eq7} %
\left(\frac{\dot{A}}{A}\right)^2 &=&
-\frac{k(r)}{A^2}+\frac{\rho_0(r)}{3A^3} \, ,\\
\label{eq:LTB rho 2} %
\rho(t,r) &=& \frac{(\rho_0 r^3)_{, r}}{3 A^2 r^2 (Ar)_{, r}} \, .
\end{eqnarray}
The solution of Eqs.\ (\ref{eq4}) and (\ref{eq5}) can now be written as
\begin{eqnarray}
\label{LTB soln2 R} a(\eta,r) &=& \frac{\rho_0(r)}{6k(r)}
     \left[ 1 - \cos \left( \sqrt{k(r)} \, \eta \right) \right] \, ,\\
\label{LTB soln2 t} t(\eta,r) &=& \frac{\rho_0(r)}{6k(r)}
     \left[ \eta -\frac{1}{\sqrt{k(r)}} \sin
     \left(\sqrt{k(r)} \, \eta \right) \right] + t_{b}(r) \, ,
\end{eqnarray}
where $\eta \equiv \tau\, r = \int^t dt'/A(t',r) \,$ and $A(t(\eta,r),r)=a(\eta,r)$.

In the rest of paper we will use this last set of equations .
Furthermore, without loss of generality, we may set 
the function $\rho_0(r)$ to be a constant,
 $\rho_0(r)=\rho_0=\mbox{constant}$, corresponding to the choice of coordinates in which $M(r)\propto r^3$, and we will call this, following \cite{Nambu:2005zn}, the FLRW gauge.

We need three functions to define a LTB solution, but because of the invariance under general coordinate transformations, only two of them are really independent. This implies that two observables are in principle sufficient to solve the inversion problem of mapping observations to a specific LTB model, for example the luminosity distance $D_L(z)$ and the redshift spherical shell mass $m(z)n(z)=mn(z)$.

As observed by \cite{Celerier:2009sv}, there has been sometime some confusion about the general type of LTB models which could be used to explain cosmological observations, so it is important to stress that without restricting the attention on models with homogeneous big bang, a void is not necessary to explain both $D_L(z)$ and $m(z)n(z)=mn(z)$ without the cosmological constant, but we will prove that this is only possible for LTB models that contain a weak central singularity.
 
%We call $mn(z)$ redshift spherical shell mass, since this quantity it is not the galaxy number counts as it is called %in the interesting paper  \cite{Celerier:2009sv}, but the product of the source number density $n(z)$ times the source %mass function $m(z)$, and it has dimension of energy.

\section {Geodesic equations}

We will adopt the same method developed in \cite{Romano:2009xw} to find the null geodesic equation in the coordinates $(\eta,t)$, but here instead of integrating numerically the differential equations we will find a local expansion of the solution around $z=0$ corresponding to the point $(t_0,0)\equiv(\eta_0,t)$, where $t_0=t(\eta_0,r)$.
We will also provide more details about the geodesic equation derivation which were presented in \cite{Romano:2009xw} in a rather concise way. 
We will indeed slightly change notation to emphasize the fully analytical r.h.s. of the equations obtained in terms of $(\eta,t)$, on the contrary of previous versions of the light geodesic equations which require some numerical calculation of $R(t,r)$ from the Einstein's equation(\ref{eq2}). 

For this reason this formulation is particularly suitable for the derivation of analytical results.

The luminosity distance for a central observer in a LTB space 
as a function of the redshift is expressed as
\be
D_L(z)=(1+z)^2 R\left(t(z),r(z)\right)
=(1+z)^2 r(z)a\left(\eta(z),r(z)\right) \,,
\ee
where $\Bigl(t(z),r(z)\Bigr)$ or $\Bigl((\eta(z),r(z)\Bigr)$
is the solution of the radial geodesic equation
as a function of the redshift.

The past-directed radial null geodesic is given by
\bea
\label{geo1}
\frac{dT(r)}{dr}=f(T(r),r) \,;
\quad
f(t,r)=\frac{-R_{,r}(t,r)}{\sqrt{1+2E(r)}} \,.
\eea
where $T(r)$ is the time coordinate along the null radial geodesic as a function of the the coordinate $r$.

From the implicit solution, we can write 
\bea
T(r)=t(U(r),r) ,\\
\frac{dT(r)}{dr}=\frac{\partial t}{\partial \eta} \frac{dU(r)}{dr}+\frac{\partial t}{\partial r} ,
\eea

where $U(r)$ is the $\eta$ coordinate along the null radial geodesic as a function of the the coordinate $r$.

Since it is easier to write down the geodesic equation in the coordinate $(t,r)$ we will start from there 
 \cite{Celerier:1999hp}:
\begin{eqnarray}
{dr\over dz}={\sqrt{1+2E(r(z))}\over {(1+z)\dot {R'}[r(z),t(z)]}} . 
\label{eq:34} \\
\nonumber
{dt\over dz}=-\,{R'[r(z),t(r)]\over {(1+z)\dot {R'}[r(z),t(z)]}} . 
\label{eq:35} \\
\end{eqnarray}

where the $'$ denotes the derivative respect to $r$ and the dot $\dot{}$ the derivative respect to $t$. 
These equations are derived from the definition of redshift and by following the evolution of a short time interval along the null geodesic $T(r)$.

The problem is that there is no exact analytical solution for $R(t,r)$, so the r.h.s. of this equations cannot be evaluated analytically but requires to find a numerical solution for $R$ first \cite{Hellaby:2009vz}
 , and then to integrate numerically the differential equation, which is a quite inconvenient and difficult numerical procedure.

Alternatively a local expansion for $R(t,r)$ around $(t_0,0)$ ,corresponding to the central observer, could be derived and used in eq.(\ref{eq:35}), but being an expansion will loose accuracy as the redshift increases. 

For this reason it is  useful for many numerical and analytical  applications to write the geodesic equations for the coordinates $(\eta,r)$,
\bea
\label{geo3}
\frac{d \eta}{dz}
&=&\frac{\partial_r t(\eta,r)-F(\eta,r)}{(1+z)\partial_{\eta}F(\eta,r)}=p(\eta,r) \,,\\
\label{geo4}
\frac{dr}{dz}
&=&-\frac{a(\eta,r)}{(1+z)\partial_{\eta}F(\eta,r)}=q(\eta,r) \,, \\
F(\eta,r)&=&-\frac{1}{\sqrt{1-k(r)r^2}}\left[\partial_r (a(\eta,r) r)
+\partial_{\eta} (a(\eta,r) r) \partial_r \eta\right]  \, , 
\eea
where $\eta=U(r(z))$ and $F(\eta,r)=f(t(\eta,r),r)$.
It is important to observe that the functions $p,q,F$ have an explicit analytical form which can be obtained from $a(\eta,r)$ and $t(\eta,r)$ as shown below.
 
The derivation of the implicit  solution $a(\eta,r)$ is based on the use of 
the conformal time variable $\eta$, which by construction 
satisfies the relation,
\be
\frac{\partial\eta(t,r)}{\partial t}=a^{-1} \,.
\ee

This means
\bea
t(\eta,r)
&=&t_b(r)+\int^{\eta}_{0}a(\eta^{'},r) d\eta^{'} \, ,
\\
dt&=&a(\eta,r)d\eta+\left(\int^{\eta}_{0}
\frac{\partial a(\eta^{'},r)}{\partial r} d\eta^{'}+t_b^{'}(r)\right) dr \,,
\eea
In order to use the analytical solution we need to find an analytical expression for $F$ and $F_{,\eta}$.

 This can always be done by using
\bea
%&&t(0,r)=t_b(r)\,, \quad a(0,r)=0 \, , \\
&& \frac{\partial}{\partial t}=a^{-1}{\frac{\partial}{\partial \eta}} \\
%&&\frac{\partial t}{\partial r}(\eta,r)=\int^{\eta}_{0}\frac{\partial a(\eta^{'},r)}{\partial r} %d\eta^{'}+t_b^{'}(r) \, , 
&&\partial_r t(\eta,r)=
\frac{ \rho_0 \, k'(r)}{12 k(r)^{5/2}} 
\left[
3 \sin{ \left( \eta\sqrt{k(r)} \right) }
-\eta \left( 2+\cos{ \left(\eta\sqrt{k(r)}\right) } \sqrt{k(r)} \right)
\right]+t_b'(r)  \, , \\
&&\partial_r \eta=
-a(\eta,r)^{-1}\partial_r t  \, 
%&&\frac{\partial^2R}{\partial t\partial r}
%=a^{-1}\left(\frac{\partial^2 R(\eta,r)}{\partial \eta \partial r}
%+\frac{\partial^2 R(\eta,r)}{\partial \eta^2}\frac{\partial \eta}{\partial r}
%+\frac{\partial R(\eta,r)}{\partial \eta}\frac{\partial}{\partial \eta}
%\left(\frac{\partial \eta}{\partial r}\right)\right) \,.
\eea 
In this way the coefficients of equations (\ref{geo3}) and (\ref{geo4}) are 
fully analytical, which is a significant improvement over previous 
approaches.

\section{Calculating $D_L(z)$ and $mn(z)$}

Expanding the r.h.s. of the geodesics equation we can easily integrate the corresponding polynomial $q(z),p(z)$, to get $r(z)$ and $\eta(z)$.
It can be easily shown that in order to obtain $D_L(z)$ to the third order and $mn(z)$ the fourth we need to expand $r(z)$ to the third order and $\eta(z)$ to the second.

In the FLRW gauge in order to to avoid a weak central singularity \cite{Vanderveld:2006rb} we need the following expansion for $k(r)$ and $t_b(r)$:
\bea
k(r)&=&k_0+k_2 r^2\\
t_b(r)&=&t^b_0+t^b_2 r^2 \,
\eea

which are based on the fact that taking only even powers the functions are analytical everywhere, including the center. 

Here we will not give the formulae in terms of $\eta_0$ and trigonometric functions, since they are rather complicated and not relevant to the scope of this paper, but rather introduce the following quantities:

\bea
a_0=a(\eta_0,0)=\frac{\tan(\frac{\sqrt{k_0}\eta_0}{2})^2 \rho_0}{3 k_0 \tan(\frac{\sqrt{k_0}\eta_0}{2})^2+3 k_0} \label{a0} \\
H_0=\frac{3 k_0^{3/2} \left(\tan(\frac{\sqrt{k_0}\eta_0}{2})^2+1\right)}{\tan(\frac{\sqrt{k_0}\eta_0}{2})^3 \rho_0} \\
q_0=\frac{1}{2} \left(\tan(\frac{\sqrt{k_0}\eta_0}{2})^2+1\right) \label{q0}
\eea

where we have used
\bea
%T=\tan(\frac{\sqrt{k_0}\eta_0}{2})\\
H_0 & = &\frac{\dot{a}(t_0,0)}{a(t_0,0)} \label{H} \\
q_0 &= -&\frac{\ddot{a}(t_0,0)\dot{a}(t_0,0)}{\dot{a}(t_0,0)^2} \label{q}
\eea

The derivative respect to $t$ is denoted with a dot, and is calculated using the analytical solution $a(\eta,r)$ and the derivative respect to $\eta$ :
\be
\dot{a}=\partial_{t}a=\partial_{\eta}a\, a^{-1}
\ee

We may actually set $a_0=1$ by choosing an appropriate system of units, but we will leave it in order to clearly show the number of independent degrees of freedom of the problem and to emphasize the difference with $\Lambda CDM$ models.

It is important to observe that in general, without imposing the smoothness conditions, we have eight independent parameters

\be
\rho_0,\eta_0,t^b_0,t^b_1,t^b_2,k_0,k_1,k_2
\ee

The conformal time coordinate of the central observer $\eta_0$ is not really independent, in the sense that it has to be consistent with the age of the universe, which, if we assume the inhomogeneities to be only local, should be approximatively the same as the one estimated in $\Lambda CDM$ models.

We get six constraints from the expansion of $mn(z)$ and $D_L(z)$ respectively to fourth an third order, so in principle without imposing the smoothness conditions $t^b_1=k_1=0$  we should be able to solve the inversion problem of locally mapping a LTB model to any given $\Lambda CDM$ model, even taking into account the fact that, as we will show later, $mn(z)$ and $D_L(z)$ don't depend on $t^b_0$.

It is also clear that a simple preliminary argument based on counting the number of independent parameters imply that the inversion problem cannot be solved for a smooth model, i.e. if $t^b_1=k_1=0$, since we have six constraints and only five truly independent parameters. 

We will show this more explicitly in the rest of the paper.

After re-expressing the results in terms of $H_0,q_0,a_0$ we get
\bea
\eta(z)& = & \eta_0 +\eta_1 z+\eta_2 z^2 \nonumber\\
r(z)& = & r_1 z+r_2 z^2+r_3 z^3  \nonumber\\
\eta_1& = &-\frac{1}{a_0 H_0}  \\
\eta_2& = &\frac{\sqrt{2 q_0-1} \left(a_0^4 H_0^4 \left(4 q_0^3-3 q_0+1\right)-2 a_0^2 H_0^3 (1-2 q_0)^2 t^b_2-3
   k_2\right)+2 k_2 (q_0+1) \arctan{\sqrt{2 q_0-1}}}{2 a_0^5 H_0^5 (2 q_0-1)^{5/2}} \nonumber \\
r_1 & = & \frac{1}{a_0 H_0} \\
r_2& = &-\frac{q_0+1}{2 a_0 H_0} \\
r_3& = &\frac{1}{2 a_0^5 H_0^5(2 q_0-1)^3} \bigg[(2 q_0-1) \left(a_0^4 H_0^4 (1-2 q_0)^2 \left(q_0^2+1\right)-2 a_0^2 H_0^3 (1-2 q_0)^2 q_0 t^b_2-5 k_2 q_0+k_2\right)+ \nonumber \\
&&+6 k_2 \sqrt{2 q_0-1} q_0^2 \arctan{\sqrt{2 q_0-1}}\bigg]  
\eea

%\bea
%r(z)& = & r_1 z+r_2 z^2+r_3 z^3+r_4 z^4 \nonumber\\
%r_1 & = & \frac{1}{H_0} \nonumber\\
%r_2 & = & -\frac{q_0+1}{2 H_0} \nonumber\\
%r_3 & = & \frac{(2 q_0-1) \left(H_0^4 (1-2 q_0)^2 \left(-\frac{2 q_0 t^b_2}{H_0}+q_0^2+1\right)-5 k_2
%   q_0+k_2\right)+6 k_2 \sqrt{2 q_0-1} q_0^2 \arccos\left(\frac{1}{\sqrt{2 q_0}}\right)}{2 H_0^5 (2
%   q_0-1)^3} \nonumber \\
%r_4 & = &\frac{H_0^4 (1-2 q_0)^2 \left(5 q_0^3-q_0^2+4\right)-2 H_0^3 (1-2 q_0)^2 q_0 (10 q_0+1) t^b_2-3 k_2
%   \left(10 q_0^2+5 q_0-2\right)}{8 H_0^5 (2 q_0-1)^2} \nonumber \\
%   & &+ \frac{8 H_0^2 q_0 (1-2 q_0)^3 t^b_3+6 k_2 q_0^2 \sqrt{2 q_0-1} (10 q_0+1) \arccos {\left(\frac{1}{\sqrt{2}
%   \sqrt{q_0}}\right)}}{8 H_0^5 (2 q_0-1)^3}
%\eea
It is important to observe that this formulae are valid, by analytical continuation, also for negative $k_0$, i.e. also for $q_0<1/2$.

We can then calculate the luminosity distance :
\bea
D_L(z)&=&(1+z)^2r(z)a(\eta(z),r(z))=D_L^1 z+D_L^2 z^2+D_L^3 z^3 \\
D_L^1&=&\frac{1}{H_0}\\
D_L^2&=&\frac{1-q_0}{2 H_0}\\
D_L^3&=&\frac{1}{2 a_0^4 H_0^5 (2 q_0-1)^{5/2}}\bigg[\sqrt{2 q_0-1} \left(a_0^4 H_0^4 (1-2 q_0)^2 (q_0-1) q_0-2 a_0^2 H_0^3 \left(4 q_0^3-3q_0+1\right) t^b_2-9 k_2 q_0\right) \nonumber \\
&&+6 k_2 q_0 (q_0+1) \arctan{\sqrt{2 q_0-1}}\bigg]   
\eea

From the definition of $mn(z)$ and the equation for the energy density we can write

\be
4\pi mn(z) dz = \rho d^3 V=\frac{8\pi M'}{\sqrt{1-k(r)r^2}}dr
\ee

from which by using $dr=(dr/dz)dz$ we get

\be
mn(z)=\frac{2 M'(r(z)}{\sqrt{1-k(r(z))r(z)^2}}\frac{dr(z)}{dz}=\frac{\rho_0 r(z)^2}{\sqrt{1-k(r(z))r(z)^2}}\frac{dr(z)}{dz}
\ee

where in the last equation we have used the FLRW gauge condition $M(r)=\rho_0 r^3/6$, which allows to calculate $mn(z)$ directly from $r(z)$.
%From the exapnsion of $mn(z)$
%\be
%mn(z)=\left(\frac{1}{4} z^4 \rho_0 \left(k_0 \text{r1}^5+10 \text{r1}^2 \text{r3}+10 \text{r1} \text{r2}^2\right)+\frac{1}{4} z^5 \text{$\rho
%   $0} \left(6 k_0 \text{r1}^4 \text{r2}+12 \text{r1}^2 \text{r4}+24 \text{r1} \text{r2} \text{r3}+4 \text{r2}^3\right)+\frac{1}{2} \text{r1}^3 z^2
%   \rho_0+2 \text{r1}^2 \text{r2} z^3 \rho_0\right)
%\ee
%we can see that we can determine the galaxy number counts up to fifth order from the fourth order expansion of $r(z)$.

We finally get:
\bea
mn(z)&=&mn_2 z^2+mn_3 z^3 +mn_a z^4 \\
mn_2 &=&\frac{6 q_0}{H_0} \\
mn_3 &=&-\frac{12 q_0 (q_0+1)}{H_0}\\
%mn_4 &=&\frac{3 \sqrt{2-\frac{1}{q_0}} q_0^{3/2} \left((2 q_0-1) \left(a_0^4 H_0^4 (1-2 q_0)^2 \left(15 q_0^2+14
%   q_0+13\right)-20 a_0^2 H_0^3 (1-2 q_0)^2 q_0 t^b_2+10 k_2 (1-5 q_0)\right)+60 k_2 \sqrt{2 q_0-1}
%   q_0^2 \arccos\left(\frac{1}{\sqrt{2 q_0}}\right)\right)} {4 a_0^4 H_0^5 (2 q_0-1)^{7/2}}
mn_4 &=&\frac{3 q_0} {2 a_0^4 H_0^5 (2 q_0-1)^3} \bigg[(2 q_0-1) (a_0^4 H_0^4 (1-2 q_0)^2 (15 q_0^2+14 q_0+13)+ \\
&&-20 a_0^2 H_0^3 (1-2
   q_0)^2 q_0 t^b_2+10 k_2 (1-5 q_0))+60 k_2 \sqrt{2 q_0-1} q_0^2 \arctan{\sqrt{2 q_0-1}}\bigg] \nonumber
\eea

%\bea
%mn(z) & = & \frac{3 q_0 z^2}{H_0}-\frac{6 q_0 (q_0+1) z^3}{H_0} \nonumber \\
%   & & +\frac{3 q_0}{4 H_0^5 (-1 + 2 q_0)^(5/2)}\bigg[\sqrt{2 q_0-1} \left(H_0^4 (1-2 q_0)^2 \left(15 q_0^2+14 q_0+13\right)+10 k_2 (1-5 q_0)\right) %\nonumber \\
% & & +20 H_0^3 (1-2 q_0)^2 q_0 \sqrt{2 q_0-1} t^b_2+60 k_2 q_0^2 \arccos\left(\frac{1}{\sqrt{2 q_0}}\right)\bigg]z^4 \nonumber \\
%  && + \bigg[\sqrt{2 q_0-1} \left(H_0^4 (1-2 q_0)^2 \left(28 q_0^3+24 q_0^2+21 q_0+19\right)-3 k_2 \left(50 q_0^2+31
%   q_0-10\right)\right) \nonumber \\
%    & & +\sqrt{2 q_0-1} \left(24 H_0^2 (1-2 q_0)^2 q_0 t^b_3-6 H_0^3 (1-2 q_0)^2 q_0 (14 q_0+5)        t^b_2\right) \nonumber \\
%  & &  +18 k_2 (14 q_0+5) q_0^2 \arccos\left(\frac{1}{\sqrt{2 q_0}}\right)\bigg]z^5
%   & & +18 k_2 q_0^2 \sqrt{2 q_0-1} (14 q_0+5) \arccos\left(\frac{1}{\sqrt{2 q_0}}\right)\left]z^5  
%\eea

As it can be seen the effects of large-scale inhomogeneities show only from the fourth order, and while also the fifth order has been calculated \cite{Romano:2009qx}
, we don't report it here since we don't need it.

\section{Relation with apparent observables}
In this section we will briefly show the results obtained so far with apparent observables:

\bea
H^{app}(z)=\left[\frac{d}{dz}\left(\frac{D_L(z)}{1+z}\right)\right]^{-1} \label{Happ}\\ 
Q^{app}(z)=\frac{d}{dz}\left(\frac{D_L(z)}{1+z}\right)=H^{-1}(z) \\
q^{app}(z)=-1-\frac{d ln(Q(z))}{d ln(1+z)} \label{qapp}
\eea

By apparent here we mean that they are obtained using the relations expressed in eq.(\ref{Happ}-\ref{qapp}), which are true only in a flat FLRW space, but assuming that the observed luminosity distance used in these formulas correspond to a LTB model. 
 
As it could be expected on the basis of simple continuity arguments, for a smooth LTB models the central values of the apparent observables coincide with their dynamical definitions from eq.(\ref{appeq1}-\ref{appeq2}) :

\bea
H_0  = \frac{\dot{a}(t_0,0)}{a(t_0,0)} = H^{app}(0) \label{appeq1}=H^{app}_0\\
q_0 = -\frac{\ddot{a}(t_0,0)\dot{a}(t_0,0)}{\dot{a}(t_0,0)^2} =q^{app}(0)=q^{app}_0 \label{appeq2}
\eea

While it is worth mentioning this relation in order to better understand the physical meaning of the quantities we have introduced, it is important to observe that eq.(\ref{a0}-\ref{q0}) can be just considered a convenient and natural re-definition of the parameters of the problem based, and that we don't't need neither use anywhere in our calculations the concept of apparent $q^{app}(z)$ and $H^{app}(z)$ introduced above.

It is also important to observe that our results are consistent with the fact that $q^{app}_0$ must be positive in absence of a weak central singularity, since the value calculated in eq.(\ref{q0}) is always positive for any value of $\eta_0$ or $k_0$.
It can be shown that equations(\ref{appeq1}-\ref{appeq2}) are not satisfied for not smooth models \cite{Romano:2009mr}, which explains why in that case the inversion problem can be solved, since in those models $q^{app}_0$ can be negative.

\section{Calculating $D_L(z)$ and $mn(z)$ for $\Lambda CDM$ models.}
The metric of a $\Lambda CDM$ model is the FLRW metric, a special case of LTB solution, where :
\bea
\rho_0(r)&=&\rho_0\\
k(r)&=&0 \\
t_b(r) &=&0
\eea

In this section we will calculate independently the expansion of the luminosity distance and the redshift spherical shell mass for the case of a flat $\Lambda CDM$.

We will also use these formulas to check the results given in the previous section, since in absence of cosmological constant and large-scale inhomogeneities they should coincide.

One of the Einstein equation can be expressed as:

\bea
H(z)&=&H_0\sqrt{\Omega_m{\left(\frac{a_0}{a}\right)}^3+\Omega_{\Lambda}}=H_0\sqrt{\Omega_m{(1+z)}^3+\Omega_{\Lambda}}
\eea

We can then calculate the luminosity distance using the following relation, which is only valid assuming flatness:

\bea
D_L(z)=(1+z)\int^z_0{\frac{d z'}{H(z')}}
\eea

From which we can get:
\bea
D_L^1&=&\frac{1}{H_0}\\
D_L^2&=&\frac{4 \Omega_{\Lambda}+\Omega_{M}}{4 H_0}\\
D_L^3&=&\frac{-10 \Omega_{\Lambda} \Omega_{M}-\Omega_{M}^2}{8 H_0}
\eea

It is convenient to re-express the above coefficient in terms of the two observables $H_0,q_0$
\bea
D_L^1&=&\frac{1}{H_0} \\
D_L^1&=&\frac{1-q_0}{2 H_0} \\
D_L^1&=&\frac{3 q_0^2+q_0-2}{6 H_0}
\eea

where we have used the following relations
\bea
\Omega_L+\Omega_M&=&1 \\
\Omega_M&=&\frac{2 q_0+2}{3}
\eea

It should be underlined here that $H_0,q_0,a_0$ which appearing in this formulas are not the same as the ones defined in the previous section for LTB models.

For calculating $mn(z)$ we first need $r(z)$, which can be obtained from the 
radial null geodesic equation which in this case take the simplified form
\be
\frac{d r}{d z}=\frac{1}{(1+z)\dot{a}}=\frac{1}{a_0 H}
\ee

which can be easily integrated to give
\bea
r(z)&=&\frac{1}{a_0}\int^z_0\frac{d z'}{H(z')}= \\
& &\frac{z}{a_0 H_0}-\frac{(q_0+1) z^2}{2 a_0 H_0}+\frac{\left(3 q_0^2+4 q_0+1\right) z^3}{6 a_0
   H_0}+O\left(z^4\right)
\eea

%As previously observed it is important to say the $a_0$ appearing in the above formula is different the one defined in %the previous section, since the two metric are different, because the FLRW corresponds to a flat space, while the LTB %depends on the function $k(r)$.

We can now calculate $mn(z)$
\bea
\rho_0&=&3 a_0^3 \Omega_M H_0^2 \\
mn(z)&=&mn_2 z^2+mn_3z^3+mn_z^4=\frac{\rho_0 r(z)^2}{\sqrt{1-k(r(z))r(z)^2}}\frac{dr(z)}{dz}=\\
&&\frac{2(q_0+1) z^2}{ H_0}-\frac{4 (q_0+1)^2 z^3}{ H_0}+\frac{5 (q_0+1)^2 (9 q_0+5) z^4}{6
   H_0}+O\left(z^5\right)
\eea

We can check the consistency between these formulae and the one derived in the case of LTB without cosmological constant by setting:
\bea
k_2=t_2^b=0\\
q_0=1/2
\eea

which corresponds the case in which $\Omega_M=1$.

It is important to mention again that $a_0,q_0,H_0$ defined in this section are in general different from the ones defined in the previous section for LTB models but for simplicity of notation we have used the same symbols.

In the next section, where we will study the inversion problem, we will instead need to clearly distinguish between them and for that reason we will introduce $a_0^{\Lambda},q_0^{\Lambda},H_0^{\Lambda}$ to indicate the ones corresponding to a $\Lambda CDM$ model.

\section{The inversion problem cannot be solved}
In this section we will denote with an upper script $\Lambda$ all the relevant quantities referred to a $\Lambda CDM$ model, including the coefficients of the redshift expansion for $D_L(z)$ and $mn(z)$.
 
In order to solve he inversion problem we need to solve the following system of six equations equations:
\bea
D_i^{\Lambda}=D_i& 1\leq i\leq 3\\
mn_j^{\Lambda}=mn_j& 2\leq j\leq 4 
\eea

It is evident that the first and second order term of the luminosity distance expansion imply that
\bea
q_0=q_0^{\Lambda}\\
H_0=H_0^{\Lambda}
\eea

If we try match the second and third order term of $mn(z)$ instead we get:
\bea
q_0=q_0^{\Lambda} \\
H_0=\frac{(3 H_0^{\Lambda} q_0^{\Lambda})}{(1 + q_0^{\Lambda})}
\eea
 
It is clear that there is no common solution unless $q_0=\frac{1}{2}$, which corresponds exactly to set to zero the cosmological constant.

This proof is general, and it does not depend on the particular value of $\Omega_{\Lambda}$, as long as it is non zero.
In this regards this is an even stronger statement than the one already well known fact that $q_0$ cannot be negative for a smooth LTB model.

In fact even if $q_0>0$, which for a flat $\Lambda CDM$ would imply $\Omega_{\Lambda}<\frac{1}{3}$, it would still be impossible to mimick the cosmological constant for both $mn(z)$ and $D_L(z)$.

In other words no value of the cosmological constant can be mimicked, even if for positive $q_0$, which is in principle possible even in a $\Lambda CDM$ models with a smaller value of $\Omega_{\Lambda}$ than the presently observed one, which is about $0.7$, and which implies a negative $q_0$.

This means that smooth LTB models are fundamentally different from $\Lambda CDM$ models, and the sign of $q_0$ it is not the really important observational feature to focus on, but we should rather consider the full observational implications at various redshift, since quantities like $mn(z)$ encode much more information about the underlying space-time geometry than simply $q_0$.

%%It is important to observe that we did not need to use higher order terms in the expansion, i.e. the ones which depends on %%the 	ities of the LTB solution.

If we had relaxed the condition $t_1^b=k_1=0$ the inversion problem could be solved, but we will give the solution in a separate future work \cite{Romano:2009mr}, where we will also try to better understand the nature of the singularity occurring in this case.

%This means that a it is intrinsically impossible to mimick the effects of a cosmological constant for both $mn(z)$ and $D_L(z)$.

From our calculations we have learnt that we can mimick the effects of a cosmological for each observable separately, but taking into consideration both of them allow to uniquely distinguish between a smooth LTB and $\Lambda CDM$ models.
In this sense since an accurate calculation of $mn(z)$ from observational data does not exist yet, there is still an open possibility that the observed $mn(z)$ is not the one corresponding to the the $\Lambda CDM$ model but it is better fitted by an appropriate LTB model, in which case $mn(z)$ would be the observable which would allow us to rule out $\Lambda CDM$ as a realistic cosmological model. 

This was actually the main motivation to introduce $E_{RSS}(z)$, to test self-consistently in redshift space the homogeneity of matter distribution.
In this regard $E_{RSS}(z)$ is actually more suitable for analyzing observational data from galaxy surveys because it can avoid the effects of astrophysical evolution of the sources.

\section{Conclusion}

We have calculated for a central observer in a LTB space without cosmological constant the low redshift expansion for  the luminosity distance and the redshift spherical shell mass respectively to third and fourth order. We have then derived the same observables for $\Lambda CDM$ models and shown how it is impossible to mimick dark energy with a smooth LTB matter dominated universe .
Our results imply that future accurate analysis of galaxy surveys data could allow us to determine $mn(z)$ to distinguish without any ambiguity between smooth $LTB$ and $\Lambda CDM$ models.
%In this way we disprove same recent claims that there exist appropriate LTB models which are undistingushable from %$\Lambda CDM$ as long as  $D_(z)$ and $mn(z)$ are concerned.

What we have shown does not diminish the importance of the study of the cosmological implications of large-scale inhomogeneities, but simply give a definite proof that smooth pressureless spherically symmetric matter inhomogeneities are not sufficient for a central observer to explain the observational data corresponding to a flat $\Lambda CDM$ model.
%As proposed in \cite{Romano:2008,Starobisky:2008}, a more promising path could actually be the study of large-scale inhomogeneities in presence of a cosmological constant, or even of a more general type of dark energy.
%It would be for example interesting to estimate how the cosmological constant value or the effective equation of state of dark energy could be affected by large-scale inhomogeneities.
%This problems will be addressed in future separate works.

\begin{acknowledgments}
I thank A. Starobinsky and M. Sasaki for useful comments and discussions, and J. Yokoyama for the 
for the hospitality at RESCUE. This work is supported by MEXT Grant-in-Aid for the global COE program
 at Kyoto University, "The Next Generation of Physics, Spun from Universality and Emergence".

\end{acknowledgments}

\end{document}